\documentstyle[sprocl]{article}
\def\be{\begin{eqnarray}}
\def\en{\end{eqnarray}}
\def\non{\nonumber}

\def\lsim{ {\ \lower-1.2pt\vbox{\hbox{\rlap{$<$}\lower5pt\vbox{\hbox{$\sim$}
}}}\ } }
\def\gsim{ {\ \lower-1.2pt\vbox{\hbox{\rlap{$>$}\lower5pt\vbox{\hbox{$\sim$}
}}}\ } }

\def\pr{{\sl Phys. Rev.}~}
\def\prl{{\sl Phys. Rev. Lett.}~}
\def\pl{{\sl Phys. Lett.}~}

\def\zp{{\sl Z. Phys.}~}

\begin{document}
\title{MESONIC FORM FACTORS IN THE LIGHT-FRONT QUARK MODEL}
\author{ HAI-YANG CHENG }
\address{Institute of Physics, Academia Sinica, Taipei, Taiwan 115, Republic
of China}
\maketitle\abstracts{
   Form factors for $P\to P$ and $P\to V$ transitions due to the
valence-quark configuration are calculated directly in the physical
time-like range of momentum transfer within the light-front quark model.
It is pointed out that the Bauer-Stech-Wirbel type of 
light-front wave function fails to give a correct normalization for the 
Isgur-Wise function at zero recoil in $P\to V$ transition. Some of the 
$P\to V$ form factors are found to depend on the recoiling direction of the 
daughter mesons relative to their parents. Thus, the inclusion of the 
non-valence contribution arising from 
quark-pair creation is mandatory in
order to ensure that the physical form factors are independent of the 
recoiling direction. }

  The $q^2$ dependence of the mesonic form factors is customarily assumed to
be governed by near pole dominance in most existing approaches. In principle,
QCD sum rules, lattice QCD simulations, and quark models allow one to 
compute form-factor
$q^2$ behavior. However, the analyses of the QCD sum rule yield some 
contradicting results. For example, while $A_1^{B\rho}$ is found to 
decrease from $q^2=0$ to $q^2=15\,{\rm GeV}^2$ in \cite{BBD}, such 
a phenomenon is not seen in \cite{ABS,YH}. Also the sum-rule results
become less reliable at large $q^2$ due to a large cancellation between
different terms. The present lattice QCD technique does not allow to 
compute the form factors directly in weak $B$ decays. Additional assumptions
on extrapolation from charm to bottom scales and from $q^2_{\rm max}$
to other $q^2$ have to be made.

   As for the quark model, the form factors evaluated in the non-relativistic
quark model is trustworthy only when the recoil momentum of the daughter
meson relative to the parent one is small. As the recoil momentum increases,
we have to start considering relativistic effects
seriously. In particular, at the maximum recoil point $q^2=0$ where the final
meson could be highly relativistic, there is no reason to expect that
the non-relativistic quark model is still applicable. A consistent treatment
of the relativistic effects of the quark motion and spin in a bound state
is a main issue of the relativistic quark model. To our knowledge,
the light-front quark model \cite{Ter} is the only relativistic quark
model in which a consistent and fully relativistic treatment of quark spins
and the center-of-mass motion can be carried out. 
In previous works \cite{Jaus,Don94a} using $q^+=0$, 
one can only calculate form 
factors at $q^2$ =0; moreover, the form factors $f_-(q^2)$ in $P\to P$ decay 
and $a_-(q^2)$ in $P\to V$ decay cannot be studied.
Hence extra assumptions are needed to 
extrapolate the form factors to cover the entire 
range of momentum transfer. Based on the dispersion
formulation, form factors at $q^2>0$ were obtained in \cite{Mel}
by performing an analytic continuation from the space-like $q^2$ region. 
Finally, the weak form factors for $P\to P$ transition were calculated
in \cite{Cheung2,Sima,Simb} for the first time for the entire range of
$q^2$, so that additional extrapolation assumptions are no
longer required. This is based on the observation \cite{Dubin} that in 
the frame where the momentum transfer is purely longitudinal, i.e., 
$q_\perp=0$, $q^2=q^+q^-$ covers the entire range of momentum transfer.
The price one has to pay is that, besides the
conventional valence-quark contribution, one must also consider the 
non-valence configuration
(or the so-called $Z$-graph) arising from quark-pair creation from the vacuum.

For the first time, we have calculated the $P\to V$ form factors directly at 
time-like momentum transfers by evaluating the form factors in a frame
where $q^+\geq0$ and $q_\perp=0$. Our main results are \cite{CCH}:

   1). We have investigated the behavior of the form factors
in the heavy-quark limit and found that the requirements of 
heavy-quark symmetry for $P\to P$ transition and 
for $P\to V$ transition are indeed fulfilled by the light-front quark model
provided that the universal function $\zeta(v\cdot v')$ obtained from
$P\to V$ decay is identical to the Isgur-Wise function $\xi(v\cdot v')$
in $P\to P$ decay.

  2). Contrary to the Isgur-Wise function in $P\to P$ decay, the 
normalization of $\zeta(v\cdot v')$ at zero recoil depends on the light-front 
wave function used. We found that the Baure-Stech-Wirbel (BSW) \cite{BSW} 
amplitude correctly gives
$\xi(1)=1$, but $\zeta(1)=0.87$. Therefore, this type of wave functions
cannot describe $P\to V$ decays in a manner consistent with heavy-quark
symmetry. The incomplete overlap of wave functions at zero recoil in $P\to V$ 
transition implies spin symmetry breaking. In other words, when the spin-1
Melosh transformation acts on the BSW wave function, it will induce a
spin-symmetry breaking term not suppressed by $1/m_Q$.

   3). Using the Gaussian-type amplitude, the Isgur-Wise function $\zeta
(v\cdot v')$ has a correct normalization at zero recoil and is identical
to $\xi(v\cdot v')$ numerically up to six digits. It can be fitted very well
with a dipole dependence with $M_{\rm pole}=6.65$ GeV for $B\to D$ transition.
 However, the predicted slope parameter $\rho^2=1.24$ is probably too large,
probably due to the lack of enough high-momentum components at large $k_\perp$
in the wave function.

    4). The valence-quark and non-valence contributions to form factors 
are in general dependent on the recoiling direction of the daughter 
meson relative to the parent meson, but their sum should not. Although we do 
not have a reliable estimate of the pair-creation effect, we have 
argued that, for heavy-to-heavy transition, form factors calculated from the 
valence-quark configuration evaluated in the
``+" frame should be reliable in a broad kinematic region, and they become
most trustworthy in the vicinity of maximum recoil.

  5). The form factors $F_1,~A_0,~A_2,~V$ (except for $V^{B\rho}$ and 
$V^{BK^*}$) all exhibit a dipole behavior, which
$F_0$ and $~A_1$ show a monopole behavior in the close vicinity of maximum
recoil for heavy-to-light transition, and in a broader kinematic region
for heavy-to-heavy decays. Therefore, $F_1,~A_0,A_2,~V$ increase
with $q^2$ faster than $F_0$ and $A_1$.

  6). In the following we present some numerical results for form factors
at maximal recoil $q^2=0$ evaluated in the ``+'' frame. We found \cite{CCH}
\be
F_1^{BD}(0)=F_0^{BD}(0)=0.70\,,
\en
and
\be
f_+^{B\pi}(0)=0.29\,,~~f_+^{BK}(0)=0.34\,,~~f_+^{D\pi}(0)=0.64\,,~~
f_+^{DK}(0)=0.75\,, 
\en
for $P\to P$ transitions, and 
\be
V^{BD^*}(0)=\,0.78\,,~A^{BD^*}_0(0)=\,0.73\,,~
A^{BD^*}_1(0)=\,0.68\,,~A^{BD^*}_2(0)=\,0.61\,, \label{BDS0}
\en
as well as
\be
&& A_0^{BK^*}(0)=0.32\,,~A_1^{BK^*}(0)=0.26\,,~A_2^{BK^*}(0)=
0.23\,,~V^{BK^*}(0)=0.35\,,   \non \\
&& A_0^{DK^*}(0)=0.71\,,~A_1^{DK^*}(0)=0.62\,,~A_2^{DK^*}
(0)=0.46\,,~V^{DK^*}(0)=0.87\,,    \non \\
&&  A_0^{B\rho}(0)=0.28\,,~~A_1^{B\rho}(0)=0.20\,,~~A_2^{B
\rho}(0)=0.18\,,~~V^{B\rho}(0)=0.30\,,    \\
&&  A_0^{D\rho}(0)=0.63\,,~~A_1^{D\rho}(0)=0.51\,,~~A_2^{D
\rho}(0)=0.34\,,~~V^{D\rho}(0)=0.78\,,   \non 
\en
for $P\to V$ transitions. 
Experimentally, only $D\to K^*$ form factors have been measured 
with the results \cite{PDG}
\be
V^{DK^*}(0)=1.1\pm 0.2\,,~~A_1^{DK^*}(0)=0.56\pm 0.04\,,~~A_2^{DK^*}(0)=
0.40\pm 0.08\,,  
\en
obtained by assuming a pole behavior for the $q^2$ dependence. Our predictions
for the $D\to K^*$ form factors are consistent with experiment.
   Two form-factor ratios defined by
\be
R_1(q^2) &=& \left[1-{q^2\over (M_B+M_{D^*})^2}\right]\,{V^{BD^*}(q^2)
\over A_1^{BD^*}(q^2)},   \non \\
R_2(q^2) &=& \left[1-{q^2\over (M_B+M_{D^*})^2}\right]\,{A_2^{BD^*}
(q^2)\over A_1^{BD^*}(q^2)},   
\en
have been extracted by CLEO \cite{R12} from an analysis of angular 
distribution in $\bar{B}\to D^*\ell\bar{\nu}$ decays with the results:
\be
R_1(q^2_{\rm max})=\,1.18\pm 0.30\pm 0.12\,,~~~~R_2(q^2_{\rm 
max})=\,0.71\pm 0.22\pm 0.07\,.
\en
Our light-front calculations yield $V^{BD^*}
(q^2_{\rm max})=1.14\,,~A^{BD^*}_1(q^2_{\rm max})=0.83\,, $ and $~A^{BD^*}_2
(q^2_{\rm max})=0.96$, hence $R_1(
q^2_{\rm max})=1.11$ and $R_2(q^2_{\rm max})=0.92\,$, 
in agreement with experiment.

\section*{Acknowledgments}
  I wish to thank C. Y. Cheung and C. W. Hwang for fruitful collaboration.
   This work was supported in part by the National Science Council of ROC
under Contract No. NSC85-2112-M-001-010.

\newcommand{\bi}{\bibitem}

\end{document}